  \documentstyle[12pt]{article}
  
\catcode`@=11
\def\secteqno{\@addtoreset{equation}{section}%
\def\theequation{\thesection.\arabic{equation}}}
\catcode`@=12
\secteqno
\newcommand{\be}{\begin{equation}}
\newcommand{\ee}{\end{equation}}
\newcommand{\bea}{\begin{eqnarray}}
\newcommand{\eea}{\end{eqnarray}}
\newcommand{\bref}[1]{(\ref{#1})}
\newcommand{\nn}{\nonumber}

\def\vsej{\vskip 4mm} 
\newcommand{\slPi}{/ {\hskip-0.27cm{\Pi}}}
\newcommand{\slP}{/ {\hskip-0.27cm{P}}}
\newcommand{\slD}{/ {\hskip-0.27cm{D}}}
\newcommand{\slX}{/ {\hskip-0.27cm{X}}}
\newcommand{\slp}{/ {\hskip-0.27cm{p}}}

\begin{document}
\vfill
\vbox{
\hfill February\ 14,\ 2000\null\par
\hfill KEK-TH-674\null\par
\hfill YITP-00-6
}\null
\vskip 20mm
\begin{center}
{\Large\bf BPS states carrying fermionic brane charges}\par
\vskip 10mm
{\large Machiko\ Hatsuda~~and~~Makoto\ Sakaguchi $^\dagger$}\par
\medskip
Theory Division,\ High Energy Accelerator Research Organization (KEK),\\
\ Tsukuba,\ Ibaraki,\ 305-0801, Japan \\
$\dagger$ Yukawa Institute for Theoretical Physics,\ Kyoto University,\\
\ Sakyou-ku,\ Kyoto,\ 606-8502, Japan\\
\medskip
mhatsuda@post.kek.jp,\ 
sakaguch@yukawa.kyoto-u.ac.jp
\medskip
\vskip 10mm
\end{center}
\vskip 10mm
\begin{abstract}
 The Green-Siegel central extension of superalgebras for BPS branes 
is studied.
 In these cases commutators of usual bosonic brane charges 
only with the broken supersymmetry charges
allow this central extension.
 We present an interpretation of these fermionic central charges 
as fermionic brane charges,
and show that they take nonzero values for a nontrivial 
fermionic boundary condition. 
 Such fermionic coordinates solutions are determined by
equations of motion in a suitable gauge condition
which manifests Nambu-Goldstone fermionic modes
as well as bosonic modes in the static gauge. 
 We also show that some modes of dilatino fields couple with 
the fermionic brane currents. 
\end{abstract} 
\par
\newpage
\setcounter{page}{1}
\parskip=7pt
\section{ Introduction}\par
\indent

For supersymmetric theories superalgebras are powerful tool to explore
non-perturbative aspects about BPS states \cite{Wit,Hull,Tow}.
About a decade ago
Green showed the possibility of the central extension of the Super-Poincar$\acute{\rm e}$
algebra \cite{Green}, where the momentum charge does not commute with the supercharge
and gives rise to the fermionic central charge\footnote{$Q,P,Z$ are a supercharge, 
a momentum and a fermionic central charge.
 $\Gamma$ and $C=i\Gamma^0$ are gamma matrix and 
a charge conjugation matrix, a slash represents contraction with $\Gamma^m$,
e.g. $\slP=P_m\Gamma^m$.}
:
\bea
\{Q_\alpha,Q_\beta\}=2(C\slP)_{\alpha\beta}~~,~~
[P_m,Q_\alpha]=(Z\Gamma_m)_\alpha ~~.\label{newsusy}
\eea
Siegel proposed a good use of this algebra
 that the Wess-Zumino action of the Green-Schwarz superstring 
can be obtained as a local bilinear form \cite{Siegel}, namely an element of the 
Chevalley-Eilenberg cohomology \cite{azcTow}.
Recently this idea has been applied to superbrane theories and super-D-brane theories
\cite{Sez,Sakag,AHKT,azc1,Sakag2}.
As well as a superstring these objects 
require the Wess-Zumino actions
to include the kappa symmetry.
The Wess-Zumino actions for $p$-branes and D-$p-$branes are higher rank tensors
and/or containing U(1) fields,
so integrating and obtaining their local expressions
are complicated \cite{KH,MK}. 
These complications can be simplified,
if Siegel's method is used.  
As long as the algebra \bref{newsusy} tells the truth in this sense,
its physical interpretation will be the next question.

It is interesting to compare the global superalgebra with the local superalgebra. 
In order to realize the global superalgebra one needs the local supersymmetry constraints $F_\alpha=0$
 to remove extra fermionic degrees of freedom.
They anticommute with the global supercharges, $\{F,Q\}=0$ \cite{WB,Siegel}.
($Q_\alpha,P_m$) are global charges, while ($F_\alpha,D_m$) 
are covariant derivatives satisfying following algebras;
\bea
~\{F_\alpha,F_\beta\}=-2(C\slD)_{\alpha\beta}~~,~~
\left[ D_m,F_\alpha \right]=(W\Gamma_m)_\alpha~~.
\eea
$W$ is named the super-Yang-Mills field strength \cite{Siegel2,WScm}.
It is also expected that $W_\alpha$ and $Z_\alpha$ are a global charge 
and a covariant derivative
respectively in a larger superspace.
In order to see the physical role of $Z_\alpha$,
it is useful to take a concrete model.
 For a superstring case charges and covariant derivatives
  are given in terms of space-time coordinates $X$ and $\theta$,
for example
\bea
W_\alpha \Leftrightarrow \bar{\theta}'_\alpha~\label{Wthe}
\eea  
where $'$ stands for the derivative with respect to worldsheet spacial coordinate $\sigma$.
\bref{Wthe} suggests that 
the new fermionic central charges are realized by 
$Z_\alpha\Leftrightarrow\int\bar{\theta}'_\alpha$.
This is a surface term, so nontrivial $Z_\alpha$ 
will require nontrivial boundary values of the fermionic coordinates.

Bergshoeff and Sezgin applied the Siegel's method to a M2 brane
where commutators of supercharges and 
not only the momentum charge but also brane charges
allow fermionic central extension  \cite{Sez}.
Further studies have been done for NS1 (IIA,IIB) and D1 \cite{Sakag}, 
for D3, D5 and NS5 \cite{Sakag2} and
for D2 and M5 \cite{azc1}.
In the work on SL(2,R)$\approx$SO(2,1) covariant central extension \cite{AHKT},
the momentum and the NS/NS and R/R brane charges are treated equally as SO(2,1)
triplet elements.
Then the SO(2,1) triplet fermionic central charges appear.
It was shown that there exists the Rarita-Schwinger type constraint on the new 
fermionic central charges and one of the triplet elements is redundant.
In other words,
the momentum or the brane charge can be chosen to be super-invariant.
In this paper we will consider the super-invariant momentum
and the super-noninvariant brane charges.
In \cite{Sez,Sakag,AHKT,azc1,Sakag2} the super-noninvariant momenta can be realized
by introducing auxiliary fields.
Instead, in this paper 
we will examine the super-noninvariant brane charges
by introducing a nontrivial fermionic boundary condition
without any auxiliary fields.

Organization of the present paper is as follows:
In section 2 we discuss the fermionic central extension of the superalgebra
 from the BPS state point of view.
The Jacobi identity of three supercharges restricts that 
the commutator of the brane charge and the broken supersymmetry charge
allows  the fermionic central extension.
In section 3 we will justify this by examining
equations of motion in a suitable gauge condition. 
We show that there exists nontrivial fermionic solution
which can give non-zero values of new fermionic brane charges. 
In section 4 we will also consider the supergravity coupling with this state.
The quantum states and the vertex operators for D-branes are not known 
because of the 
difficulty of the soliton quantization.
On the other hand classical soliton property may suggest 
how to couple with the supergravity fields.
Since it is supposed that D-branes are static objects,
the nonrelativistic approximation
will be a good approximation.
So we begin by the classical Gauss law equation,
then examine its consistency under the broken supersymmetry.
The unbroken supersymmetry transformation of 
a purely bosonic soliton solution has been studied in \cite{Dabh}.
Usually, including this reference, the fermionic supergravity fields
are set to be zero.
However we rather consider nontrivial fermionic supergravity fields.
It turns out that there can be nonzero dilatino fields whose some mode 
couple to the new fermionic brane currents.

\par
\vsej
\section{BPS state with new fermionic brane charges}
\indent

We begin by a D-string case \cite{HK} as a simplest example.
The D-string superalgebra is given by\footnote{Notation follows \cite{HK} where $(X,\theta)$
are space-time coordinates and their conjugates are $(p,\zeta)$. }
\bea
\{Q_{A\alpha },Q_{B\beta }\}&=&-2(C\Xi_G)_{A\alpha~B\beta}~~,~~
\Xi_G=\Gamma^m_{\alpha\beta}~
\left({\bf 1}_{AB} P_m-(\tau_I)_{AB}\Sigma_m^I \right)~~.\label{BPS}
\eea 
Concrete expression of the global supercharges, 
the total momenta and the brane charges are 
\bea
Q_{A\alpha}&=&\int d\sigma ~[\zeta-\bar{\theta}\slp+\bar{\theta}\slX'\tau_q
-\frac{1}{6}\bar{\theta}\Gamma\tau_q \theta'\cdot \bar{\theta}\Gamma
-\frac{1}{6}\bar{\theta}\Gamma \theta'\cdot \bar{\theta}\Gamma\tau_q
]_{A\alpha}~~,\nn\\
P_m&=&\int d\sigma ~p_m ~~,\\
\Sigma^I_m&=&q^I\int^{\sigma_F}_{\sigma_I} d\sigma ~X'_m = q^I (X_F-X_I)_m~~,
\eea
where $\sigma_{F,I}$ are D-string end points. 
The massiveness of a D-string is reflected to its infinity length of
$\Sigma\sim L |_{L \to \infty}$.

One may notice that there is no way to centrally extend of $[P,Q]$,
\bea
\left[ P_m,Q\right]&=&\left[\int d\sigma_1~ p_m(\sigma_1),\int d\sigma_2~  \bar{\theta}\slX'(\sigma_2)
\tau_q\right]\nn\\
&=&-\int d\sigma_2 ~\partial_2
\left(\int d\sigma_1 \delta(\sigma_1-\sigma_2) \right)\bar{\theta}(\sigma_2)\Gamma_m\tau_q=0
\eea
with $\tau_q=q^I \tau_I=q^{NS} \tau_3+q^{R}\tau_1$ (I=NS,R or 3,1).
On the other hand, the brane charges $\Sigma^I$  allow to have the central extension in a commutator 
with the supercharges:
\bea
\left[ \Sigma^I_m,Q\right]&=&-q^I\int d\sigma_1~ \partial_1 
\left( \int d\sigma_2  \bar{\theta}(\sigma_2)\Gamma_m \delta(\sigma_1-\sigma_2) \right) \nn\\
&=& -Z^I\Gamma_m~~.\label{SQZ}
\eea
Now $Z$ may be interpreted as a fermionic brane charge 
\bea
Z^I=q^I\int^{\sigma_F}_{\sigma_I} d\sigma~ \bar{\theta}'= q^I (\bar{\theta}_F-\bar{\theta}_I)~~.
\eea
This charge will be infinity volume
$Z\sim L \bar{\eta}|_{L \to \infty}$ with a constant spinor $\eta$, 
as same as a usual bosonic brane charge because the supertranslation does not change
the volume in general.

The superalgebra \bref{BPS} shows that a D-string ground state is BPS saturated 
where 1/2 supersymmetry are broken.
The right hand side of the \bref{BPS} is a projection operator 
\bea
C\Xi_G=-2iM{\cal P}_-~~,~~M=T_q~ L \label{Proj}
\eea
with a D-string mass $M$, a D-string tension $T_q=\sqrt{(q^I)^2}$ and a string length $L$,
and
supersymmetry(SUSY) charges are projected into unbroken SUSY charges 
(${\cal Q}$) and broken SUSY charges (${\cal S}$)  
\bea
Q&=&\left\{\begin{array}{ccc}
{\cal Q}&\equiv&Q{\cal P}_+\\
{\cal S}&\equiv&Q{\cal P}_-\end{array}\right.
\eea
in such a way that
\bea
\{{\cal Q},{\cal Q}\}=0=\{{\cal Q},{\cal S}\}~,~\{{\cal S},{\cal S}\}\neq 0
\label{QQS}
\eea
then
\bea
 {\cal Q}|0\rangle=0~,~{\cal S}|0\rangle \neq 0~,\eea
for a ground state $|0\rangle$.
The projection operators ${\cal P}_\pm$ defined as \bref{Proj} 
 become simple form\footnote{Gamma matrices is denoted as 
 $\Gamma^{m_1\cdot\cdot\cdot m_N}=$
$ \frac{1}{N!}\displaystyle{\sum_{\rm antisymmetrized\ m}}$
$ \Gamma^{m_1}\cdot\cdot\cdot \Gamma^{m_N}$. } 
 for a ground state in the static gauge
\bea
{\cal P}_\pm=\frac{1}{2}\left({\bf 1}_{\alpha\beta}{\bf 1}_{AB}
\pm (\Gamma_{01})_{\alpha\beta}(\hat{\tau}_q)_{AB} \right)~~,~~
\hat{\tau}_q=\tau_q/|\tau_q|.\label{projpm}
\eea
It is shown that the central extension \bref{SQZ} is possible only for
broken SUSY 
\bea
\left[ \Sigma_m,{\cal S}_\alpha\right]=(Z \Gamma_m)_\alpha~~,
\eea
since
\bea
\left[ \Sigma_m,{\cal Q}_\alpha\right]=0 
\eea
from the Jacobi identity of three $Q$'s by using with \bref{QQS}
\bea
\left[\{Q_{\alpha },Q_{\beta }\},{\cal Q}_{\gamma}\right]
&=&(C\Gamma^m\tau_I)_{\alpha\beta}\left[\Sigma_m^I,{\cal Q}_{\gamma}\right]
\nn\\
&=&-\left[\{{\cal Q}_{\gamma},Q_{( \alpha }\},Q_{\beta ) }\right]=0
\eea
with $A_{(\alpha}B_{\beta)}=A_{\alpha}B_{\beta}+A_{\beta}B_{\alpha}$.

The algebra discussed above is summarized as follows:
\bea
\begin{array}{lclc}
\{{\cal Q},{\cal Q}\}=\{{\cal Q},{\cal S}\}=0&,&
\{{\cal S},{\cal S}\}=-2C\Xi_G &,\\
\left[ \Sigma^I_m,{\cal Q} \right]=0&,&
\left[ \Sigma^I_m,{\cal S} \right]=-Z^I\Gamma_m
\label{algebra}&,\\
\left[ P_m,{\cal Q} \right] =0&,&\left[ P_m,{\cal S} \right] =
0&.
\end{array}\label{BPSsal}
\eea 

\vsej
\section{Gauge fixing and the equation of motion}
\indent

We examine the equation of motion 
in a suitable gauge condition and will find the ground state solution to give 
the algebra \bref{BPSsal}.
In addition to the static gauge, we impose the following gauge condition
\bea
\theta_+\equiv {\cal P}_+\theta=0 \label{gaugeth}
\eea
by using the kappa invariance.
The local superalgebra for a D-string is given by \cite{HK}
\bea
\{F_{A\alpha }(\sigma_1),F_{B\beta }(\sigma_2)\}&=&
2(C\Xi)_{A\alpha~B\beta} (\sigma_1)\delta (\sigma_1-\sigma_2)~
~,\label{FF}
\\~~
(\Xi)_{A\alpha~B\beta}&=&\Gamma^m_{\alpha\beta}~
\left({\bf 1}_{AB} \tilde{p}_m-(\tau_q)_{AB}(\Pi_1)_m \right)~~,\nn
\eea
where
\bea
F&=&\zeta+\bar{\theta}(\tilde{p}-\slPi_1 \tau_q)
-\frac{1}{2}(\bar{\theta}\Gamma \theta'\cdot \bar{\theta}\Gamma \tau_q 
+\bar{\theta}\Gamma \tau_q \theta'\cdot \bar{\theta}\Gamma )~~,\nn\\
\tilde{p}_m&=&p_m+\bar{\theta}\Gamma_m \tau_q\theta'~~,\nn\\
(\Pi_1)_m&=&X'_m-\bar{\theta}\Gamma_m\theta'~~.
\eea 
$\Xi$ is a rank half and nilpotent operator so that a half of $F$
are first class constraints $ \tilde{F}_{A\alpha}\equiv (F\Xi)_{A\alpha}=0$ 
generating the kappa symmetry and another half are second class. 
The counting of the physical degrees of freedom is following:
The degrees of freedom of $\theta^{A\alpha}$ and $\zeta_{A\alpha}$ are 
$16\times 2\times 2$.
The number of the second class constraints is $16$.
The number of the first class constraints is $16$,
and the one for the gauge fixing conditions is $16$.
So totally surviving degrees of freedom are
$16\times 2\times 2-(16+16\times 2)=16$, so 
$8\ \theta$'s and $8\ \zeta $'s are physically dynamical variables.

For a ground state in the static gauge, 
the right hand side of the \bref{FF} becomes 
the same form of the unbroken SUSY projection operator
 $C\Xi=-2iM{\cal P}_-$ of \bref{projpm}.
This coincidence occurs to guarantee the universal property, $\{Q,F\}=0$.
In order to solve BPS fermionic solutions explicitly,
we restrict ourself to discuss only on the ground state
in the rest of this section. 
It should be noticed that the kappa projection is not
$C\Xi$ but $\Xi$.
The kappa symmetry is generated by
\bea
\int \tilde{F}\kappa'=\int F\Xi\kappa'
=\int F{\cal P}_+(2M \Gamma^0 \kappa')=\int F{\cal P}_+\kappa_+ ~~,
\eea
therefore the kappa parameter is projected along the unbroken SUSY direction. 
The gauge condition \bref{gaugeth} can be possible by using the kappa transformation 
with $\delta_\kappa \theta_+=\kappa_+=-\theta_+$.

Under the global translation, $X_m$ is transformed as $\delta_a X_m=a_m$ .
Analogously under the global SUSY, $\theta_-$ is transformed $\delta_\epsilon \theta_-=\epsilon_-$,
while $\theta_+=0$ is preserved by the kappa transformation 
$(\delta_\epsilon +\delta_\kappa)\theta_+=\epsilon_++\kappa_+$ with $\kappa_+=-\epsilon_+$. 
In this gauge $X^m$ is transformed under the global SUSY
and the kappa transformation
\bea
(\delta_\epsilon +\delta_\kappa) X^m&=&\bar{\epsilon}_-\Gamma^m \theta_-
+\bar{\epsilon}_+\Gamma^m \theta_-
-\bar{\kappa}_+\Gamma^m \theta_- =\bar{\epsilon}_-\Gamma^m \theta_-
+2\bar{\epsilon}_+\Gamma^m \theta_-~~,
\eea
where $\epsilon_-$ is nothing but the broken SUSY parameter.
Of course the static gauge is always recovered by the reparametrization invariance. 
The transverse coordinates $X^i$ and $\theta_-$ are Nambu-Goldstone modes associated with
the broken translation invariance and the broken supertranslation by a brane
\cite{HPol}.
 
The action for a D-$p$-brane is
\bea
I&=&- \int d^{p+1}\sigma~ L~=~- \int d^{p+1}\sigma \left[ T\sqrt{-\det(G_{\mu\nu}+{\cal F}_{\mu\nu})} 
+{\cal L}_{WZ}\right]~~\\
G_{\mu\nu}&=&\Pi_\mu^m \Pi_{\nu m}~~,~~
\Pi_\mu^m=\partial_\mu X^m-\bar{\theta}\Gamma^m\partial_\mu \theta
~~,~~{\cal F}_{\mu\nu}=F_{\mu\nu}-\Omega^{3}_{[\mu\nu]}\nn\\
\Omega_{\mu\nu}^I&=&-(\bar{\theta}\Gamma^m\tau^I \partial_{\mu}\theta)
(\Pi_{\nu m}+\frac{1}{2}\bar{\theta}\Gamma_m\partial_\nu\theta)~~~~(I=1,3).\nn
\eea
In the fermionic gauge ${\cal P}_+\theta=\theta_+=0$ and the static gauge $X^\mu=\sigma^\mu$, 
${\mu}=0,1,\cdot\cdot\cdot ,p$, it becomes
\bea
G_{\mu\nu}&=&\eta_{\mu\nu}-\bar{\theta}_-\Gamma_{(\mu}\partial_{\nu)}\theta_-
+(\bar{\theta}_-\Gamma^{\rho}\partial_{\mu}\theta_-
)(\bar{\theta}_-\Gamma_{\rho}\partial_{\nu}\theta_-)
+\partial_\mu \phi^i \partial_\nu \phi^i \\
q^I\Omega_{\mu\nu}^I&=&-\bar{\theta}_-\Gamma_{\nu}\tau_q \partial_{\mu}\theta_-
+\frac{1}{2}(\bar{\theta}_-\Gamma^{\rho}\tau_q \partial_{\mu}\theta_-
)(\bar{\theta}_-\Gamma_{\rho}\partial_{\nu}\theta_-)\nn
\eea
where $\phi^i$, ${i}=p+1,\cdot\cdot\cdot,9$ are perpendicular modes to a $p$-brane. 
It is used that $[{\cal P}_\pm,C\Gamma^\mu]=0$,
 ${\cal P}_\pm (C\Gamma^i)=(C\Gamma^i){\cal P}_\mp$
 and 
$[{\cal P}_\pm,\tau_q]=0$.
The expression of the Dirac-Born-Infeld part is available for all $p$.
In this section we take D1 case as an example where the Wess-Zumino part is given by
\bea
{\cal L}_{WZ}|_{p=1}&=&T\epsilon^{\mu\nu}\Omega^{1}_{\mu\nu}~~,
\eea
and we will discuss general D-$p$-brane cases in the last section.
Under an arbitrary variation of $\delta \theta_-$ it becomes
\bea
\delta L&=&-\frac{TG}{2\sqrt{-\det(G+{\cal F})}}G^{\mu\nu}\delta G_{\mu\nu}
+q^I\epsilon^{\mu\nu} \delta \Omega^I_{\mu\nu}
\label{eqm}\\
\delta G_{\mu\nu}&=&-(\delta \bar{\theta}_-\Gamma_{(\mu}\partial_{\nu)} \theta_-
-\partial_{(\mu}\delta \bar{\theta}_-\Gamma_{\nu)} \theta_-)\nn\\
&&+2(\delta \bar{\theta}_-\Gamma^\rho\partial_\mu \theta_-)
 ( \bar{\theta}_-\Gamma_\rho\partial_\nu \theta_-)
-2(\partial_\mu\delta \bar{\theta}_-\Gamma^\rho \theta_-)
 ( \bar{\theta}_-\Gamma_\rho\partial_\nu \theta_-)
 \nn\\
 q^I\delta {\Omega}^I_{\mu\nu}&=&-(\delta \bar{\theta}_-\Gamma_\nu\tau_q\partial_\mu \theta_-
-\partial_\mu\delta \bar{\theta}_-\Gamma_\nu \tau_q\theta_-)
\nn\\
&&+ \frac{1}{2}\{
(\delta \bar{\theta}_-\Gamma^\rho\tau_q\partial_\mu \theta_-)
 ( \bar{\theta}_-\Gamma_\rho\partial_\nu \theta_-)
-(\partial_\mu\delta \bar{\theta}_-\Gamma^\rho \tau_q\theta_-)
 ( \bar{\theta}_-\Gamma_\rho\partial_\nu \theta_-)\nn\\
&&
 -(\delta \bar{\theta}_-\Gamma^\rho\partial_\mu \theta_-)
 ( \bar{\theta}_-\Gamma_\rho\tau_q\partial_\nu \theta_-)
+(\partial_\mu\delta \bar{\theta}_-\Gamma^\rho \theta_-)
 ( \bar{\theta}_-\Gamma_\rho\tau_q\partial_\nu \theta_-)\}
\nn ~~.
\eea 
For a ground state where $G_{\mu\nu}$ and ${\cal F}_{\mu\nu}$ are constant
and terms in $O(\theta^3)$ vanishes,
the equation of motion $\delta L/\delta\bar{\theta}_-=0$ becomes quite simple
\bea
\Gamma_\mu(\eta^{\mu\nu}-\hat{\tau}_q\epsilon^{\mu\nu})\partial_\nu 
\theta_-=0~~\label{eqmg}
\eea
where $q^I$ are given by
\bea
\left\{
\begin{array}{lcl}
q^{NS}&=&q^3=E^1=T\frac{{\cal F}_{01}}{\sqrt{-\det (G+{\cal F})}}\\
q^{R}&=&q^1=T
\end{array}\right.
\eea
and we use $(q^I)^2=T^2G/\det(G+{\cal F})$.
If we take the following ground state ansatz
\bea
\theta_-=\sigma \Theta\label{tsT}~
\eea
with a constant spinor $\Theta$, then the equation of motion \bref{eqmg} reduces into 
\bea
(\Gamma_1-\hat{\tau}_q \Gamma_0)\partial_1 \theta_-
&=&\Gamma_1(1+\hat{\tau}_q \Gamma_{01})\Theta=0~~.
\eea
Then there exists nontrivial ground state solution of $\theta_-$ as the form of
\bref{tsT} with
\bea
{\cal P}_+\Theta=0~~.
\eea
The boundary term of the equation motion also allows the solution \bref{tsT}
\bea
\Gamma_\mu(\eta^{\mu\nu}-\hat{\tau}_q\epsilon^{\mu\nu})
\theta_-|_{\rm boundary}=0~~.\label{bdc}
\eea
As a result there can be a fermionic brane charge for this solution
\bea
Z^I_\alpha=q^I\int_{\sigma_I}^{\sigma_F}~d\sigma \bar{\theta}'_{-\alpha}=q^IL~\bar{\Theta}_\alpha|_{L\to\infty}~~.
\eea
It is natural to set the volume of $Z$ to be infinity, since it is obtained from 
infinity volume brane charge $\Sigma$ by the supertransformation.

The choice $\Theta=0$ or $\theta_-=$const. brings back to the usual Super-Poincar$\acute{\rm e}$ algebra
which is usually considered as a ground state solution.

 \par
\vsej
\section{Supergravity coupling}
\indent

A D-string is described by the NS/NS and R/R worldvolume currents:
\bea
(J^{I})^{nl}(x)&=&q^{I}\int d\tau d\sigma ~\partial_\tau X^{[n}\partial_\sigma X^{l]}
\delta^{(10)}(x-X(\tau,\sigma))~~.\label{current}
\eea
In this section we take 
a ``nonrelativistic gauge" $X^0=\tau$ ($\partial_\tau X^0=1$) \cite{HYAT}
rather than the static gauge, in order to keep a spacial rotational symmetry.
The static property, equally nonrelativistic property, of D-branes is assumed.
For these objects the spacial covariance will be a suitable guiding principle
to determine the background coupling.
To preserve this gauge,
additional restriction is imposed on $\theta_-$:
\bea
\delta_{\epsilon_-}(\partial_\tau X^0-1)=
\bar{\epsilon}_-\Gamma^0 \partial_\tau \theta_-=0~~\Rightarrow~~
\partial_\tau \theta_-=0~~.\label{dtt0}
\eea
The worldvolume currents \bref{current} are related to the brane charge as
\bea
\int d^9 x(J^I)^{0m}(x)
&=&\int d^9x ~q^I \int d\tau d\sigma ~\partial_\tau X^0 \partial_\sigma X^m~
\delta^{(9)}(x-X)\delta(x^0-\tau)\nn\\
&=&q^I\int d\sigma~ \partial_\sigma X^m\nn\\
&=&(\Sigma^I)^m(\tau)\label{JandS}
\eea
in the nonrelativistic gauge.


The currents \bref{current} are sources of
 two form gauge fields $B^I_{mn}$.
 In a flat background if we put a test D-brane, 
 field equations for $B^I_{mn}$ become
\bea
\partial^l(H^I)_{lmn}(x)&=&(J^I)_{mn}(x)~~\label{dHJ}
\eea
where $H^I_{lmn}$ are fluctuations.

Now let us consider the broken supersymmetry transformation of
\bref{dHJ}.
Because the brane charge $(\Sigma^I)^m$ transforms into the fermionic brane charge 
$(Z^I)^{\alpha}$ under the broken supersymmetry and because of \bref{JandS}, 
a bosonic brane current $(J^I)^{mn}$ also transforms into 
a fermionic brane current $({\cal J}^I)^{m\alpha}$ :
\bea
\delta_{\epsilon_-} (J^I)^{mn}(x)&=&
q^I\int d\tau d\sigma \left(\partial_\tau \delta_{\epsilon_-} X^{[m} 
\partial_\sigma X^{n]}+\partial_\tau X^{[m}
\partial_\sigma \delta_{\epsilon_-} X^{n]}
\right) \delta^{(10)}(x-X)\nn\\
&=&-q^I  \int d\tau d\sigma~
 \partial_\tau X^{[m}
\partial_\sigma \bar{\theta}_-
\Gamma^{n]} {\epsilon}_- \delta^{(10)}(x-X)
\nn\\
&\equiv&-({\cal J}^I)^{[m|\alpha|}(x)(\Gamma^{n]}{\epsilon}_-)_{\alpha}~~.\label{deJ}
\eea
It is noted that 
under the broken supersymmetry
 brane coordinates transform into 
$\delta_{\epsilon_-} X^m=\bar{\epsilon}_-\Gamma^m\theta_-=-\bar{\theta}_-\Gamma^m\epsilon_-$.
The fermionic currents ${\cal J}^{m\alpha}$ 
\bea
({\cal J}^I)^{n\alpha}(x)
=q^I \int d\tau d\sigma~
 \partial_\tau X^n
\partial_\sigma \bar{\theta}_-^{\alpha}
 \delta^{(10)}(x-X(\tau,\sigma))
\eea
are also conserved currents
\bea
\partial_n({\cal J}^I)^{n\alpha}(x)
&=&q^I \int d\tau d\sigma~
 \partial_\tau X^n
\partial_\sigma \bar{\theta}_-^{\alpha}
 \partial_n\delta^{(10)}(x-X(\tau,\sigma))\nn\\
&=&-q^I \int d\tau d\sigma~
 \partial_\sigma \bar{\theta}_-^{\alpha}
(\frac{\partial X^n}{\partial \tau}
 \frac{\partial}{\partial X^n})\delta^{(10)}(x-X(\tau,\sigma))\nn\\
&=&q^I \int d\tau d\sigma~
\partial_ \tau \partial_\sigma \bar{\theta}_-^{\alpha}
\delta^{(10)}(x-X(\tau,\sigma))=0~~,
\eea
and are related with the fermionic charges $(Z^I)^{\alpha}$ as
\bea
\int d^{9}x ({\cal J}^I)^{0\alpha}(x)&=&
q^I\int d\sigma~\partial_\sigma \bar{\theta}_-^\alpha\nn\\
&=&(Z^I)^{\alpha}(\tau)~~.
\eea

If the right hand side of the equation \bref{dHJ} transforms under the 
broken supersymmetry
of the brane,
the left hand side should also transform.
In another word
the total supersymmetry charge is sum of the supergravity part and
the brane part.
Transformation rules of supergravity fields are given in \cite{SUGRAIIB},
and
\bea
\delta_\epsilon (B^I)_{mn}
=(2\bar{\psi}_{[m}\Gamma_{n]}-\bar{\lambda}\Gamma_{mn})\tau^I{\epsilon},
\eea
where $\psi$ is a gravitino and $\lambda$ is a dilatino.
In the first order of this test brane perturbation
the background metric is not affected,
 since the energy momentum tensor which is a source of the metric
  contains square of $H$ as the fluctuation.
So gravitinos must be set to zero $\psi_m=0$
because $\delta_\epsilon e_m^a\sim \bar{\psi}_m\Gamma^a \epsilon=0$.
In order to have nontrivial transformation rule of $B^I_{mn}$
with the broken SUSY parameter $\epsilon_-$,
we will consider nonzero $\lambda_+$ so that 
$\delta B_{01}=-\bar{\lambda}_+\Gamma_{01}\tau \epsilon_-\neq 0$
as their fluctuations.

Under the broken global SUSY
a dilatino 
does not transform
  $\delta_{\epsilon} \lambda \sim (\partial \phi, H)\cdot\Gamma\epsilon=0$
  in a flat background $(\partial \phi=0, H=0)$.
Then the left hand side of \bref{dHJ} transforms into
\bea
\delta_{\epsilon_-} \partial^l (H^I)_{lmn}&=&\Box \delta_{\epsilon_-}  (B^I)_{mn}
\nn\\
&=&-\frac{1}{2}(\Box\bar{\lambda}_+\Gamma_{[m}\tau^I)\Gamma_{n]}\epsilon_-\label{boxBJ}
\eea
in the Lorentz gauge $\partial^m B_{mn}=0$.
As a result of \bref{deJ} and \bref{boxBJ},
under the broken SUSY \bref{dHJ} transforms into 
\bea
\frac{1}{2}\left(\Box\bar{\lambda}_+\tau^I\Gamma^{n}\right)^\alpha(x)=
({\cal J}^I)^{n\alpha}\label{boxlJ}(x)~~.
\eea
Therefore
this new fermionic brane charge becomes a source of dilatino fields.

It is also confirmed that another current does not follow.
Even if one more broken supersymmetry is performed on the fermionic brane current,
it vanishes by \bref{dtt0} 
\bea
\delta_{\epsilon_-}({\cal J}^I)^{n\alpha}
=q^I \int 
 \partial_\tau \bar{\theta}_-^{\beta}
\partial_\sigma \bar{\theta}_-^{\alpha} 
~\delta^{(10)}(x-X)~(\Gamma^n{\epsilon}_-)_\beta 
=0~~.
\eea
This concludes that a D-string action allows 
the fermionic brane currents coupled with some modes of dilatino fields
as well as the usual brane current coupling $B_{mn}J^{mn}$,
when a D-string has a nontrivial fermionic boundary condition.

\par


\section{Discussions}\par
\indent

Generalization to arbitrary D-$p$-branes is straightforward.
The global supersymmetry and the local supersymmetry algebras for D-$p$-branes 
are given in \cite{KH,MK},
as the same form as \bref{BPS} and \bref{FF}.
The projection \bref{projpm} has the form of 
\bea
{\cal P}_\pm=\frac{1}{2}({\bf 1}\pm \hat{\Gamma})~~,\label{prjpp}
\eea 
and $\hat{\Gamma}$ is replaced by
suitable $p$-brane projection $\Gamma_{01\cdot\cdot\cdot p}\hat{\tau}_{p}$.
For example $\hat{\Gamma}=\Gamma_0\Gamma_{11}$ for D0,
$\hat{\Gamma}=\Gamma_0\Gamma_1\Gamma_2$ for D2 and so on. 
By using with these projection operators,
global supersymmetry charges are separated into 
broken supersymmetries and unbroken supersymmetries.
Breaking 16 supersymmetry, $ {\cal S}|0\rangle \neq 0$, gives rise to 
16 Nambu-Goldstone fermions $\theta_-={\cal P}_-\theta$.
Half of this, i.e. 8 $\theta_-$'s are dynamical and 
the rest are non-dynamical by the equations of motion 
in the excited states.   
Another set of 16 fermionic coordinates are gauged away  $\theta_+={\cal P}_+\theta=0$
by using the kappa symmetry generated by the first class constraints $\tilde{F}=0$. 

Equations of motions can be analyzed analogously.
In the same static gauge fixing condition ${\cal P}_+\theta=0$ of \bref{gaugeth} with
\bref{prjpp} 
and $X^\mu=\sigma^\mu$, equations of motions are given as
\bea
\Gamma_\mu\left(\eta^{\mu\nu}+
\epsilon^{\mu\rho_1\cdot\cdot\cdot\rho_{p-1}\nu}\Gamma_{\rho_1\cdot\cdot\cdot\rho_{p-1}} 
(\hat{\tau}_{\bf p})/(p-1)!\right)
\partial_\nu \theta_-=0~~,
\eea 
and the nontrivial ground state solution of \bref{tsT} can be extended as
\bea
\theta_-=\sum_l\sigma_l \Theta_l ~~(l=1,\cdot\cdot\cdot,p) ~~{\rm with} ~~
{\cal P}_+\Theta_l=0~~.\label{grdtp}
\eea 
For D-$p$-brane cases, brane charges are $p$-rank tensors.
Each commutator with the broken supersymmetry charges
replaces a vector index of the brane charges by a spinor index.
When $\theta_-$ has nontrivial solution of \bref{grdtp},
totally $p+1$ kinds of brane charges with vector indices or spinor indices   
have nonzero values.
For example a D2 brane carries totally $3$ kinds of brane charges,
 among which
$2$ kinds of bosonic brane charges $\Sigma_{mn}$ and $Z_{\alpha\beta}$ are 
obtained in the static gauge as
\bea
\Sigma_{12}=TL^2~~,~~
Z_{\alpha\beta}=TL^2\bar{\Theta}_{1(\alpha}~\bar{\Theta}_{2\beta)}~~,
\eea
and $1$ kind of fermionic brane charges $Z_{m\alpha}$ are
obtained as
\bea
Z_{1\alpha}=TL^2\bar{\Theta}_{2\alpha}~~,~~
Z_{2\alpha}=-TL^2\bar{\Theta}_{1\alpha}~~.
\eea
The supergravity coupling will follow the argument of the section 4
replacing $B_{mn}$ by 
a $p+1$-rank gauge field $C_{m_1\cdot\cdot\cdot m_{p+1}}$ for a D-$p$-brane respectively.

It is curious that considering that
the equation \bref{boxlJ} 
is second order although usual 
equation of motion for a spinor field is first order. 
Even in a flat background this fermionic brane current will couple to 
some mode of dilatino. 
So the coupling will be with
the dilaton potential mode  $(\slD /\Box) {\lambda}$
or nonlocal.
Further studies are necessary to clarify these points
and to apply many other systems.

\vskip 6mm
{\bf Acknowledgments}\par
M.H. wishes to thank Joaquim Gomis, Ken-ji Hamada, Nobuyuki Ishibashi
and Shunya Mizoguchi for
fruitful discussions,
and M.S. would like to express his gratitude to Hiroshi Kunitomo 
and wishes to thank the theory group of KEK for the kind hospitality.
We also thank Kiyoshi Kamimura for helpful discussions,
and Mitsuko Abe and Nathan Berkovits for a question ``what is $Z_\alpha$?"
which is
our motivation of this work.

\end{document}